\documentclass[11pt]{article}
\usepackage{graphics}
\usepackage{psfrag}
\usepackage{epsfig}
\usepackage{psfig}
\usepackage{epsf}
\usepackage{float}
\usepackage{amssymb,stmaryrd,latexsym}
\newcommand{\fet}[1]{\mbox{\boldmath $#1$}}

\newcommand{\beq}{\begin{equation}}
\newcommand{\eeq}{\end{equation}}
\newcommand{\beqa}{\begin{eqnarray}}
\newcommand{\eeqa}{\end{eqnarray}}
\newcommand{\nn}{\nonumber \\ }

\newcommand{\vs}{\vspace{-0.2cm}}

\begin{document}


\hfill {\tiny JLAB-THY-04-244}

\vspace{-0.4cm}

\hfill {\tiny HISKP-TH-04/14}

\vspace{-0.4cm}

\hfill {\tiny FZJ-IKP-TH-2004-09}


\vspace{1cm}

\begin{center}

{{\Large\bf Isospin dependence of the three--nucleon force
}}\footnote{Work 
supported in part by U.S.~Department of Energy under contract number DE-AC05-84ER40150.}

\end{center}

\vspace{.3in}

\begin{center}

{\large 
E. Epelbaum,$^\ast$\footnote{email: epelbaum@jlab.org}
Ulf-G. Mei{\ss}ner,$^\star$$^\ddagger$\footnote{email: 
                           meissner@itkp.uni-bonn.de}
J.E. Palomar$^\dagger$\footnote{email:
                           palomar@condor3.ific.uv.es}}

\bigskip

$^\ast${\it Jefferson Laboratory, Theory Division, Newport News, VA 23606, USA}

\bigskip

$^\star${\it Universit\"at Bonn, Helmholtz-Institut f{\"u}r
  Strahlen- und Kernphysik (Theorie)\\ D-53115 Bonn, Germany}\\

\bigskip

$^\ddagger${\it Forschungszentrum J\"ulich, Institut f\"ur Kernphysik 
(Theorie)\\ D-52425 J\"ulich, Germany}

\bigskip

$^\dagger${\it Departamento de Fisica Teorica and IFIC, Universidad 
de Valencia, \\ Institutos de Investigacion de Paterna,
46071 Valencia, Spain}

\end{center}

\vspace{.6in}

\thispagestyle{empty} 

\begin{abstract}
\noindent 
We classify $A$--nucleon forces according to their isospin dependence and 
discuss the most general isospin structure of the three--nucleon force.
We derive the leading and subleading isospin--breaking corrections to the 
three--nucleon force using the framework of chiral effective field theory.
\end{abstract}

\vfill

\pagebreak

\section{Introduction}
\def\theequation{\arabic{section}.\arabic{equation}}
\label{sec:intro}

Three--nucleon forces (3NFs) are well established in nuclear physics. Although
small compared to the dominant two--nucleon force (2NF), they are nevertheless needed
to gain a quantitative understanding of nuclei and nuclear physics. A recent example
in this context is the  discussion of the 3NF effects in proton--deuteron scattering,
see e.g.~\cite{cad00,er03}. Other examples are the binding energy difference
between $^3$H and $^3$He or the saturation properties of nuclear matter.
Only in the last decade a theoretical tool has become available to systematically
analyze few--nucleon forces and consider such fine but important aspects as
isospin--violation in such forces and in systems made of a few nucleons. This
tool is the extension and application of chiral perturbation theory to systems
with more than one nucleon which require an additional non--perturbative resummation
to deal with the shallow nuclear bound states and large S-wave scattering lengths.
While 3NFs in the isospin limit have been analyzed in some detail, see e.g.
\cite{wein,ko94,FHK,E02}, the question of isospin--violation in the 3NF has not
yet been addressed in this framework. The work reported here is intended to fill
this gap. 

Without further ado, let us address the issues considered here. First, we generalize
the classification of the isospin dependence of two--nucleon forces due to Henley
and Miller \cite{henl} to the case of $A$ nucleons ($A\ge 3$), with particular
emphasis on the three--nucleon system, see Section~\ref{sec:gen}. This is essentially
a quantum--mechanical exercise and does not reveal any of the underlying dynamics.
The keywords here are isospin mixing, charge--independence (breaking) and charge--symmetry
(breaking). We stress that while such language, which precedes QCD and originates from
Heisenberg's definition of isospin to account for the almost degeneracy of the
proton and the neutron combined with almost equality of their strong forces, 
is useful to categorize few--nucleon forces, in QCD the underlying 
broken symmetry is isospin of the light up and down quarks. This symmetry is broken
in pure QCD by the light quark mass difference and further by electromagnetism when
external electroweak interactions are considered. Thus, in the second part of this work,
Section~\ref{sec:EFT}, we derive the leading and next--to--leading order isospin--violating
contributions of the 3NF based on chiral effective field theory (EFT)\footnote{We eschew
here pionless nuclear EFT as it is not the appropriate tool to analyze this particular
problem.}. We briefly recall the counting rules for the inclusion of strong and
electromagnetic isospin violation presented in \cite{N3LO} 
and discuss the pertinent terms of the effective chiral Lagrangian in
Section~\ref{sec:pc}. We present the leading and subleading isospin--breaking
contributions to the 3NF in momentum space in Section~\ref{sec:3Nmom}, followed by a brief
estimate of the relative strength of these forces in Section~\ref{sec:est}.  
We end with a short summary. The appendix contains the coordinate space
representation of the isospin--violating 3NF. 

\section{General considerations}
\def\theequation{\arabic{section}.\arabic{equation}}
\setcounter{equation}{0}
\label{sec:gen}

This section deals with a novel classification scheme for the isospin dependence of the
$A$--nucleon forces. To derive this scheme, one does not make any assumption about the dynamics underlying
such forces but only utilizes their transformation properties under isospin symmetry
and charge symmetry operations on the level of nucleons. 

\vfill

\subsection{Definitions and notation}
\label{sec:GenRem}
The non--relativistic A--nucleon system is described by the Hamilton operator $H$ 
\beq
H = H_0 +  V^{\rm 2N} +  V^{\rm 3N} + \;\ldots  \; +  V^{\rm AN} \;,
\eeq
where $H_0$ is the nucleon kinetic energy and $V^{\rm n N}$ represents 
the potential corresponding to the $n$--nucleon force.
The total isospin operator $\fet T$ is given by the sum of the isospin operators 
$\fet t$ of the individual nucleons:
\beq
\fet T = \sum_{a=1}^A \fet{t} (a) \;.
\eeq
The total isospin operator $\fet T$ as well as the operators $\fet t (i)$ satisfy the Lie algebra
of the $SU(2)$ isospin group:
\beqa
\left[ T_i , \; T_j \right] &=& i \epsilon_{ijk} \; T_k \;, \\
\left[ t_i (a) , \; t_j (b) \right] &=& i \delta_{ab} \epsilon_{ijk} \; t_k (a) \;,  \nonumber
\eeqa
with $i,j,k = 1,2,3$.
The single--nucleon isospin operators $t_i (a)$ can be conveniently represented in terms of
Pauli matrices $\tau_i$:
\beq
t_i (a) = \frac{1}{2} \tau_i (a) \;.
\eeq

The charge operator $Q$ is defined for the A--nucleon system as:
\beq
Q = e \left( \frac{A}{2} + T_3 \right)\;.
\eeq
Since the baryon number and the charge are conserved in nuclear reactions,  the 
operator $T_3$ commutes with $H$ even if isospin symmetry is broken.

Charge symmetry represents invariance under reflection about the 1--2 plane in charge space.
The charge symmetry operator $P_{cs}$ transforms proton and neutron states into each other and 
is given by \cite{henl}:
\beq
\label{pcs}
P_{cs} = e^{i \pi T_2} = \prod_{a=1}^A e^{i \pi t_2 (a)} = \prod_{a=1}^A \left( i \tau_2 (a) \right) \;.
\eeq
Thus charge symmetry conservation means the equivalence of {\it nn} and {\it pp}, 
{\it nnn} and {\it ppp}, $\dots$, forces.
Obviously, charge symmetry  is valid if isospin is conserved, i.e. if 
\beq
\label{comminv}
\big[ H, \; \fet T^2 \big] =  \big[ H, \; T_i \big] = 0\;. 
\eeq

\subsection{Two nucleons}

The classification of the two--nucleon forces according to their 
isospin dependence has been worked out by Henley and Miller \cite{henl}. 
For the sake of completeness, we will briefly remind the reader of this 
classification scheme in what follows.

The two--nucleon forces fall into four classes:
\begin{itemize}
\item
Class (I) forces $V_{I}^{\rm 2N}$ are isospin invariant and can be expressed as:
\beq
V_{I}^{\rm 2N} = \alpha_1  + \alpha_2 \, \fet t(1) \cdot \fet t(2)\;,
\eeq
where $\alpha_i$ are space and spin operators.
\item
Class (II) forces, $V_{II}^{\rm 2N}$, maintain charge symmetry but break charge independence (i.e. are not isospin 
invariant\footnote{Clearly, $V_{II}^{\rm 2N}$ as well as all other considered isospin--violating interactions 
still commute with the third components of the total isospin for the reason explained before.}):
\beqa
\label{class2comm}
\big[ V_{II}^{\rm 2N}, \; \fet T \big] &\neq& 0 \;,  \nn
\big[  V_{II}^{\rm 2N}, \; P_{cs} \big] &=& 0 \;. 
\eeqa
The class (II) forces are proportional to the isotensor:
\beq
\label{isoten}
V_{II}^{\rm 2N} = \alpha \, \tau_3 (1) \, \tau_3 (2) \;.
\eeq
It is easy to verify that these forces do not mix isospin in the two--nucleon system and thus satisfy,
in addition, the following relation:
\beq
\label{class2addit}
\big[ V_{II}^{\rm 2N}, \; \fet T^2 \big] = 0 \;.  
\eeq
\item
Class (III) forces break charge symmetry but do  not lead to isospin mixing in the two--nucleon system:
\beqa
\big[ V_{III}^{\rm 2N}, \; \fet T \big] &\neq& 0 \;,  \nonumber \\
\big[  V_{III}^{\rm 2N}, \; P_{cs} \big] &\neq& 0 \;, \nn
\big[ V_{III}^{\rm 2N}, \; \fet T^2 \big] &=& 0 \,. 
\eeqa
Such forces have the general structure:
\beq
V_{III}^{\rm 2N} = \alpha ( \tau_3 (1) + \tau_3 (2) ) \;.
\eeq
and are symmetric under the interchange of the nucleons 1 and 2.
\item
Finally, class (IV) forces break charge symmetry and cause isospin mixing, i. e.:
\beqa
\big[ V_{IV}^{\rm 2N}, \; \fet T \big] &\neq& 0 \;,  \nonumber \\
\big[  V_{IV}^{\rm 2N}, \; P_{cs} \big] &\neq& 0 \;, \nn
\big[ V_{IV}^{\rm 2N}, \; \fet T^2 \big] &\neq& 0 \,.
\eeqa
They can be expressed as:
\beq
V_{IV}^{\rm 2N} = \alpha_1 ( \tau_3 (1) - \tau_3 (2) ) + \alpha_2 [ \fet \tau (1) \times \fet \tau (2) ]_3 \;.
\eeq
The operator $\alpha_2$ has to be odd under a time reversal transformation.
\end{itemize}

\subsection{Three and more nucleons}
\label{sec:3N}

Let us now generalize the above treatment to systems with more than two nucleons. 
Considering the commutation relations of the Hamilton operator $H$ with the operators $\fet T^2$ and $P_{cs}$, 
one can distinguish between four different cases for isospin--violating forces: the Hamilton operator may commute 
with both $\fet T^2$ and $P_{cs}$, with one of those operators or with none.\footnote{In case of two 
nucleons, only three of 
these four cases appear, since there are no forces which commute with $P_{cs}$ and do not with the operator 
$\fet T^2$.}
The problem with such a classification scheme is that conservation of $\fet T^2$ depends on the number of particles. 
In general, an $A$--nucleon force that commutes with the squared total isospin operator in the $A$--nucleon system,
$\fet T_A^{\, 2} \equiv (\fet t (1) + \fet t(2) + \ldots + \fet t(A) )^2$, will not commute with the 
operator $\fet T_{>A}^{\, 2}$. For example, all isospin--breaking two--nucleon forces, which do not 
cause isospin mixing in the two--nucleon system, lead to isospin mixing in the 
three--nucleon system.  
On the other hand, the property of charge symmetry is independent on the number of nucleons and suitable for 
generalization. 
Thus in systems with more then two nucleons it is convenient to distinguish 
between the following three classes of forces:
class (I) isospin symmetric forces, class (II) forces, which break isospin but maintain charge 
symmetry and class (III) forces, which break both isospin and charge symmetry. For two nucleons,
our class (III) interactions obviously include the class (III) and (IV) forces in the classification
by Henley and Miller.

Let us now concentrate on the 3N force and list all possible isospin structures.
\begin{itemize}
\item
Class (I) forces are isospin scalars and have the structure:
\beq
V_{I}^{\rm 3N} = \sum_{i \neq j \neq k} \left( \alpha_{I}^{ijk}  +  \beta^{ijk}_{I} \, 
\fet \tau (i) \cdot \fet \tau (j) +  
\gamma^{ijk}_{I} \left[ \fet \tau (i) \times \fet 
\tau (j) \right] \cdot \fet \tau (k) \right)\,,
\eeq
where $\alpha_{I}^{ijk}$, $\beta_{I}^{ijk}$ and $\gamma_{I}^{ijk}$ are space and spin 
operators with the superscripts being the nucleon labels. 
\item
Class (II) forces satisfy:
\beqa
\big[ V_{II}, \; \fet T \big] &\neq& 0 \;,  \\
\big[  V_{II}, \; P_{cs} \big] &=& 0 \;, \nonumber
\eeqa
and can be expressed as
\beq
\label{class23n}
V_{II}^{\rm 3N} = \sum_{i \neq j \neq k} \left( \alpha^{ijk}_{II} \, t_3 (i) t_3(j)  + \beta^{ijk}_{II} 
\, [ \fet \tau (i) \times \fet \tau (j) ]_3 \tau_3 (k) \right) \;. 
\eeq
The forces in eq.~(\ref{class23n}) give rise to isospin mixing in the 3N system except in the 
following two cases:
\beqa
\label{2cases}
\alpha^{123}_{II} + \alpha^{213}_{II} &=& \alpha^{132}_{II} + \alpha^{312}_{II} = 
\alpha^{231}_{II} + \alpha^{321}_{II}\;, \nn
\beta^{123}_{II} - \beta^{213}_{II}&=& \beta^{312}_{II} -\beta^{132}_{II} = \beta^{231}_{II} 
- \beta^{321}_{II}\;. 
\eeqa 
\item
Class (III) forces satisfy:
\beqa
\big[ V_{II}, \; \fet T \big] &\neq& 0 \;,  \\
\big[  V_{II}, \; P_{cs} \big] &\neq& 0 \;. \nonumber
\eeqa
There are four types of such isospin--breaking forces:
\beqa
\label{class33n}
V_{III}^{\rm 3N} 
&=& \sum_{i \neq j \neq k} \left( \alpha_{III}^{ijk} \, \tau_3 (i)  + \beta^{ijk}_{III} \, 
[ \fet \tau (i) \times \fet \tau (j) ]_3  
+ \gamma^{ijk}_{III} \, \tau_3 (i) \, \fet \tau (j) \cdot \fet \tau(k)  \right.\nn
&& \mbox{\hskip 1.4 true cm} \left. +  \kappa_{III}^{ijk} \, \tau_3 (i) \, \tau_3 (j) \, \tau_3 (k) \right) \;. 
\eeqa
The first three terms in eq.~(\ref{class33n})  cause 
isospin mixing in the 3N system except in the following special cases:
\beqa
&& \alpha^{123}_{III} + \alpha^{132}_{III} = \alpha^{213}_{III} + \alpha^{231}_{III}= 
\alpha^{312}_{III} + \alpha^{321}_{III} \;, \nonumber \\ 
&& \beta^{123}_{III} - \beta^{213}_{III} =  \beta^{312}_{III} -\beta^{132}_{III} = 
\beta^{231}_{III} -  \beta^{321}_{III} \;, \nn
&& \gamma^{123}_{III} + \gamma^{132}_{III} =  \gamma^{213}_{III} + \gamma^{231}_{III}= \gamma^{312}_{III} + 
\gamma^{321}_{III}\;. 
\eeqa
The last term in eq.~(\ref{class33n}) does not lead to isospin mixing in the 3N system.
Notice further that the quantities $\beta_{ijk}$ are time--reversal--odd. 
\end{itemize}

In what follows, we will perform explicit calculation of the dominant isospin--violating
three--nucleon forces based on chiral effective field theory.

\section{Isospin--breaking three--nucleon force in chiral effective field theory}
\def\theequation{\arabic{section}.\arabic{equation}}
\setcounter{equation}{0}
\label{sec:EFT}

\subsection{Power counting and effective Lagrangian}
\label{sec:pc}

Isospin--breaking two--nucleon forces have been extensively studied within effective field
theory approaches, see e.g.~\cite{kolck,vKNij,kolck96,Ep99,WME,friar03,friar04}, as well as using more
phenomenological methods, see e.g.~\cite{coon96,nis02} for some recent references. 
In  the Standard Model,  isospin--violating effects have their origin 
in both strong (i.e. due to the different masses of the up and down quarks) 
and electromagnetic interactions (due to different charges of the up and down quarks).
The electromagnetic effects can be separated
into the ones due to soft and hard photons. While effects of hard photons 
are incorporated in effective field theory  
by inclusion of electromagnetic short distance  operators in the 
effective Lagrangian, soft photons have to be taken into account explicitly.

\medskip\noindent
Consider first isospin breaking in the strong interaction. The QCD quark mass term can be expressed as 
\begin{equation}
\label{eq1}
\mathcal{L}_{\rm mass}^{\rm QCD} = -\bar{q} \, \mathcal{M} \, q = -\frac{1}{2}\bar{q} \, 
(m_{\rm u}+m_{\rm d})(1-\epsilon\tau_{3})\,q~,
\end{equation}
where 
\beq\label{epsdef}
\epsilon \equiv {m_d-m_u \over m_d+m_u} \sim {1 \over 3}~.
\eeq 
The above numerical estimation is based on the light quark mass values
utilizing a modified  $\overline{\rm MS}$ subtraction scheme
at a renormalization scale of 1~GeV \cite{Le96}. 
The isoscalar term in eq.~(\ref{eq1}) breaks chiral but 
preserves isospin symmetry. It leads to the nonvanishing pion mass, 
$M_\pi^2 = (m_u + m_d ) B \neq 0$, 
where $B$  is a low--energy constant (LEC) that describes the strength of the  bilinear light quark condensates. 
All chiral--symmetry--breaking interactions in the effective Lagrangian are proportional to positive powers of $M_\pi^2$.
The isovector term ($\propto \tau_3$) in eq.~(\ref{eq1}) breaks isospin symmetry and generates 
a series of isospin--breaking effective interactions $\propto (\epsilon  M_\pi^2)^n$
 with $n \geq 1$.
It therefore appears to be natural to count strong isospin violation in terms of $\epsilon  M_\pi^2$. 
However, we note already here that isospin--breaking effects are
in general much smaller than indicated by the numerical value of $\epsilon$, because the
relevant scale for the isospin--conserving contributions is the chiral--symmetry--breaking scale 
$\Lambda_\chi$ rather than $m_u+m_d$.  

\medskip\noindent
Electromagnetic terms in the effective Lagrangian can be generated using the method of external sources, 
see e.g. \cite{urech95,MS,MM} for more details. All such terms are proportional to the nucleon charge matrix 
$Q= e \, (1 + \tau_3 )/2$, where $e$ denotes the electric charge.\footnote{Or equivalently, one can use 
the quark charge matrix $e \, (1/3 + \tau_3 )/2$.}
More precisely, the vertices which contain (do not contain) the photon fields are proportional to $Q^{n}$
($Q^{2n}$), where $n=1,2,\ldots$. Since we are interested here in nucleon--nucleon scattering in the absence of 
external fields, so that no photon can leave a Feynman diagram, it is convenient to introduce the 
small parameter $e^2 \sim 1/10$ for isospin--violating effects caused by the electromagnetic interactions. 
As will be discussed below, three--nucleon forces due to virtual photon exchange do not contribute at 
the leading and subleading orders. We will therefore not consider virtual photons in the present work. Notice however 
that electromagnetic effects might be enhanced at low energy due to the long range of the corresponding interaction, 
see \cite{N3LO} for more details. A systematic study of such effects should therefore be performed in the future. 
For the first step in this direction see \cite{Kiev03}.

\medskip\noindent
In the present study we adopt the same power counting rules for isospin--breaking contributions as introduced 
in \cite{N3LO}. Specifically, we count 
\beq\label{CountRules1}
\epsilon \sim e \sim \frac{q}{\Lambda}\,,
\eeq
where $q \sim M_\pi$ refers to a generic low--momentum scale and $\Lambda$ to the hard scale which enters the 
values of the corresponding low--energy constants. 
In addition, we keep track of the additional factors $1/(4 \pi )^2$ arising from the photon loops by 
counting 
\beq\label{CountRules2}
\frac{e^2}{(4 \pi )^2}  \sim \frac{q^4}{\Lambda^4}\,.
\eeq
Notice further that contrary to the standard practice in the single--nucleon sector, the nucleon mass is 
considered as a much larger scale compared to the chiral--symmetry--breaking scale for reasons 
explained in \cite{wein}. In this work we adopt the counting rule $q/m \sim 
(q/ \Lambda )^2$, which has also been used in \cite{ubi}. 
Counting the nucleon mass in this way ensures that all iterations of the leading--order NN potential 
contribute to the scattering amplitude at leading order $(q/\Lambda)^0$ and thus have to be resummed.
The N--nucleon force receives contributions of the order $\sim (q/\Lambda )^\nu$,
where 
\beq
\label{powc}
\nu = -4 + 2 n_\gamma + 2 N + 2 L + \sum_i V_i \Delta_i\,.
\eeq
Here, $L$ and $V_i$ refer to the number of loops and vertices of type $i$ 
and $n_\gamma$ is the number of virtual photons.
Further, the vertex dimension  $\Delta_i$ is given by
\begin{equation}
\label{chirdim}
\Delta_i = d_i + \frac{1}{2} n_i - 2\;,
\end{equation}
where  $n_i$ is the number of nucleon field operators and $d_i$ is the $q$--power of the 
vertex, which accounts for the number of derivatives and insertions of pion mass, $\epsilon$ and
$e/(4 \pi )$ according to eqs.~(\ref{CountRules1}), (\ref{CountRules2}). 

\medskip\noindent
Let us now specify the relevant terms in the effective Lagrangian. 
In the purely pionic sector, we have to take into account the following structures:
\beq
\label{pipi}
\mathcal{L}_{\pi\pi} = \frac{F_\pi^2}{4} \langle u_\mu u^\mu + \chi_+ \rangle + C \langle Q_+^2 - Q_-^2 \rangle \,,
\eeq
where $F_\pi$ refers to the  pion decay constant and the brackets $\langle \, \, \rangle$ 
denote traces in the flavor space. We remark that various LECs appearing in the effective 
Lagrangian correspond to bare quantities in the chiral SU(2) limit. Throughout this manuscript we 
will not specify this explicitly and  use physical
values for the LECs to express our results for the 3NF. 
Mass and coupling constant renormalization is detailed e.g. in refs.~\cite{BKM95,EMG03}. 
Further, 
\beqa
&& u_\mu = i ( u^\dagger \partial_\mu u 
- u \partial_\mu u^\dagger ) \,, \quad \quad u = \sqrt{U}\,, \quad \quad 
\chi = 2 B \mathcal{M}\,,  \nn
&& \chi_\pm = u^\dagger \chi u^\dagger \pm u \chi^\dagger u\,, \quad \quad \quad
Q_\pm = \frac{1}{2} (u^\dagger Q u \pm u Q u^\dagger )\,, \quad \quad
\eeqa
The unitary  2 $\times$ 2 matrix $U$ in the flavor space collects 
the pion fields. In the $\sigma$--model gauge, it takes the form 
\beq
\label{sigmodg}
U = \frac{1}{F_\pi} \left[ \sqrt{F_\pi^2 - \fet{\pi}^2} 
+ i \fet \tau \cdot \fet \pi  \right]\,.
\eeq 
The pion mass resulting from eq.~(\ref{pipi}) is given by
\beqa
M_{\pi^0}^2 &=& B (m_u + m_d)\,, \nn
M_{\pi^\pm}^2 &=& B   (m_u + m_d) + \frac{2}{F_\pi^2} e^2 C\,.
\eeqa
The experimentally known pion mass difference $M_{\pi^\pm} - M_{\pi^0} = 4.6$ MeV allows 
to fix the value of the LEC $C$, $C = 5.9 \cdot 10^{-5}$ GeV$^4$. Notice that the natural scale 
for this LEC is $F_\pi^2 \Lambda^2 /(4 \pi)^2 \sim 3 \cdot 10^{-5}$ GeV$^4$ if one adopts $\Lambda \sim M_\rho$.

Utilizing the heavy baryon framework, the relevant structures in the single--nucleon Lagrangian are \cite{fet01} 
(for a more detailed discussion see e.g the review \cite{BKMrev}):
\beqa
\label{piN}
\mathcal{L}_{\pi N} &=& \bar N_v \Big[ \, i v \cdot D + g_A \, S \cdot u  \nn
&& \mbox{\hskip 1 true cm}
+ c_1 \langle \chi_+ \rangle + \frac{c_3}{2} \langle u \cdot u \rangle + \frac{c_4}{2}  [ S^\mu , \; S^\nu ] [u_\mu , \; u_\nu ]
+ c_5 \hat \chi_+ \nn
&&  \mbox{\hskip 1 true cm}
+ f_1 F_\pi^2 \langle Q_+^2 - Q_-^2 \rangle + f_2 F_\pi^2 \langle Q_+ \rangle Q_+ + f_3 F_\pi^2 \langle Q_+ \rangle^2 \Big] N_v\,,
\eeqa
where $N_v$ refers to the field operator of a nucleon moving with the velocity $v_\mu$, $c_{1,3,4,5}$, $f_{1,2,3}$ 
are the strong and the electromagnetic LECs, respectively,  and
\beq
D_\mu = \partial_\mu + \Gamma_\mu\,,  \quad \quad  
\Gamma_\mu = \frac{1}{2} [ u^\dagger , \, \partial_\mu u ]\,, \quad \quad
\hat \chi_+ = \chi_+ - \frac{1}{2} \langle \chi_+ \rangle\,,\quad \quad
S_\mu= \frac{1}{2} i \gamma_5 \sigma_{\mu \nu} v^\nu \,.
\eeq
Keeping the terms with at most two pion fields and switching to the nucleon rest--frame
system, the Lagrangian density in eq.~(\ref{piN}) can be expressed in a more convenient form:\footnote{Notice that 
only terms with three and more pion fields depend on the specific parametrization of the matrix $U$.}
\beqa
\label{piN_2}
\mathcal{L}_{\pi N} &=& N^\dagger  \Big[  i \partial_0 - \Delta m +
\frac{g_A}{2 F_\pi} \fet \tau \vec \sigma \cdot \vec \nabla \fet \pi 
- \frac{1}{4 F_\pi^2} \fet \tau \cdot ( \fet \pi \times \dot{\fet \pi } ) \nn
&& - \frac{2 c_1}{F_\pi^2} M_\pi^2 \fet \pi^2 + \frac{c_3}{F_\pi^2} (\partial_\mu \fet \pi \cdot \partial^\mu \fet \pi ) - 
 \frac{c_4}{2 F_\pi^2} \epsilon_{ijk} \, \epsilon_{abc} \, \sigma_i \tau_a (\nabla_j \, \pi_b ) (\nabla_k \, \pi_c )  
-\frac{c_5}{F_\pi^2} \epsilon M_\pi^2 (\fet \pi \cdot \fet \tau ) \pi_3 \nn
&& + f_1 \, e^2 (\pi_3^2  - \fet \pi^2 ) + \frac{1}{4} f_2 \, e^2 ( (\fet \pi \cdot \fet \tau ) \pi_3 - \fet \pi^2 \tau_3 ) \Big] N
\eeqa
Notice that at the order we are working, there is no need to distinguish between $M_{\pi^0}$ and  $M_{\pi^\pm}$
in eq.~(\ref{piN_2}). We have therefore used the same symbol $M_\pi$ for both charged and neutral pion masses. 
The nucleon mass shift $\Delta m$ in the above equation is given by 
\beq 
\label{m_shift}
\Delta m = - 4 c_1 M_\pi^2 - \frac{1}{2} F_\pi^2 e^2 ( 2 f_1 + f_2 + 2 f_3 ) - \frac{1}{2} \tau_3 
(4 c_5 \epsilon M_\pi^2 + f_2 \, e^2 F_\pi^2 ) \,.
\eeq
The isospin invariant shift given by the first two terms in eq.~(\ref{m_shift}) is of no importance 
and can be absorbed by a redefinition of the bare nucleon mass. The proton--to--neutron mass difference 
$\delta m \equiv m_p - m_n$ fixes 
the values of the LECs $c_5$ and $f_2$ through
\beqa
\label{m_shift1}
(\delta m )^{\rm str.} &\equiv& (m_p - m_n )^{\rm str.} = -4 c_5 \epsilon M_\pi^2 
= (-2.05 \pm 0.3) \mbox{ MeV ,} \nn
(\delta m )^{\rm em.} &\equiv& (m_p - m_n )^{\rm em.} = - f_2 \, e^2 F_\pi^2 
= (0.7 \pm 0.3)  \mbox{ MeV .}
\eeqa
These values are taken from \cite{Ga82}. The electromagnetic shift is based on an evaluation of the
Cottingham sum rule. In principle, this contribution could also be evaluated in chiral perturbation
theory including virtual photons. While the formalism exists (see e.g \cite{urech95,MS,MM}), there
are still some subtleties to be addressed \cite{GSR}. Therefore, we consider the electromagnetic mass
shifts for the ground state baryon octet collected in \cite{Ga82} the best values available. 
Notice that according to the counting rules (\ref{CountRules1}) and (\ref{CountRules2}),
the strong and electromagnetic shifts in eq.~(\ref{m_shift1}) are 
effects of order $q^3$ and $q^4$, respectively. 
While the constants $c_5$ and $f_2$ can be fixed from  eq.~(\ref{m_shift1}), 
the value of the LEC $f_1$, which contribute
to isospin--violating $\pi \pi NN$ vertex, see eq.~(\ref{piN_2}), is unknown. This term plays an important
role in the analysis of isospin violation in pion--nucleon scattering and the evaluation of the 
ground state characteristics of pionic hydrogen, see \cite{fet01} and \cite{Bern}, respectively. In the
two--nucleon sector, it only leads to an isospin--invariant contribution to the TPEP at NNLO, which
has so far not been considered (it can be absorbed in the normalization of the term $\sim c_1$).
On the contrary, the resulting contribution to the 3NF is isospin--breaking. It, however, does not violate 
charge symmetry and, therefore, does e.g. not contribute to the binding--energy difference of $^3$H and $^3$He.
We further stress that the $f_1$--term  has not been included in the Lagrangian 
used in \cite{ko00,nis02,Gar04} since another power counting for the
electromagnetic effects was employed (see the discussion in \cite{kolck}).

In the few--nucleon sector we only need the following isospin invariant structures
\beqa
\label{NN}
\mathcal{L}_{NN} &=& - \frac{1}{2} C_S ( \bar N_v N_v  )  ( \bar N_v N_v  )  + 2 C_T  ( \bar N_v S_\mu N_v  )  
( \bar  N_v S^\mu N_v  ) \nn
&& - \frac{1}{2} D  ( \bar N_v N_v  )  ( \bar N_v  S \cdot u N_v  ) 
- \frac{1}{2} E  ( \bar N_v N_v  )   ( \bar N_v \fet \tau N_v  ) \cdot  ( \bar N_v \fet \tau N_v  ) \,.
\eeqa
where $C_{S,T}$, $D$ and $E$ are the corresponding low--energy constants.
The Lagrangian density (\ref{NN}) gives rise to the following relevant terms in the nucleon rest--frame system:
\beqa
\label{NN2}
\mathcal{L}_{NN} &=& - \frac{1}{2} C_S ( N^\dagger  N  )  ( N^\dagger  N  )  - \frac{1}{2} C_T  
( N^\dagger \vec \sigma N )  
( N^\dagger \vec \sigma N ) \nn
&& - \frac{D}{4 F_\pi}   ( N^\dagger N  )  ( N^\dagger  \vec \sigma \fet \tau  N  ) \cdot \vec \nabla \fet \pi 
- \frac{1}{2} E \,  ( N^\dagger N  )   ( N^\dagger \fet \tau N  ) \cdot  (  N^\dagger \fet \tau N  ) \,.
\eeqa

\subsection{Three--nucleon force in momentum space}
\label{sec:3Nmom}

\begin{figure*}[htb]
\vspace{0.5cm}
\centerline{
\psfrag{x11}{\raisebox{-0.0cm}{\hskip 0.0 true cm  (a)}}
\psfrag{x22}{\raisebox{-0.0cm}{\hskip 0.0 true cm  (b)}}
\psfrag{x33}{\raisebox{-0.0cm}{\hskip 0.0 true cm  (c)}}
\psfig{file=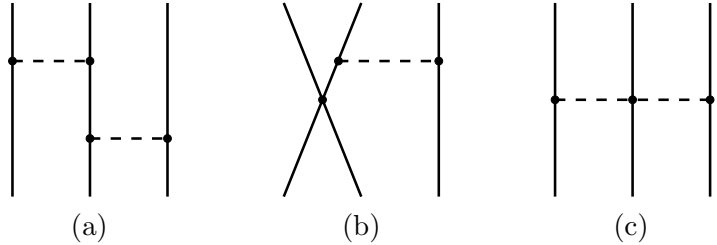,width=10cm}}
\vspace{0.3cm}
\centerline{\parbox{14cm}{
\caption[fig1]{\label{fig1} Leading contribution to the 3NF at the order $(q/\Lambda )^2$ which vanish, 
as discussed in the text. Solid and dashed lines are nucleons and pions, respectively. 
Heavy dots denote the leading--order vertices with $\Delta_i = 0$.  
}}}
\vspace{0.5cm}
\end{figure*}

We are now in the position to discuss the leading and subleading isospin--breaking contributions 
to the 3NF~\footnote{After submission of our manuscript, related work on charge symmetry
breaking in the 3N system by Friar, Payne and van Kolck \cite{FPvK}
appeared.}. 
For the sake of completeness, we will briefly remind the reader of the structure of the isospin--conserving 3NF. 
The leading 3NF contribution of the order $(q/\Lambda )^2$ represented by the graphs in Fig.~\ref{fig1} 
is well known to vanish. More precisely, the first two graphs (a) and (b) in this figure vanish in the static 
limit if one adopts an energy--independent formalism such as the method 
of unitary transformation \cite{eden96}. Alternatively, one can use old--fashioned perturbation theory to 
derive a corresponding energy--dependent 3NF potential. The latter is known to cancel against the recoil corrections 
to the 2N potential being 
iterated in the scattering equation \cite{yang86,ko94}. It should be understood that the first two diagrams 
shown in Fig.~\ref{fig1}  only specify the topology and do not correspond to Feynman graphs. Clearly, 
the corresponding contributions to the 3NF do not include the pieces generated by the iteration of the 2NF.   
We remind the reader that the operators associated with these diagrams depend on the scheme and on the definition 
of the potential. 
In the method of unitary transformation, these graphs subsume both irreducible and reducible time--ordered 
topologies. The reducible diagrams do, however, not contain anomalously small energy denominators, which correspond to 
the purely two--nucleon intermediate states in old--fashioned perturbation theory. 
The last diagram in Fig.~\ref{fig1}  is suppressed by a factor of $q/m$
due to the time derivative entering the Weinberg--Tomozawa vertex in eq.~(\ref{piN_2}).

\begin{figure*}[htb]
\vspace{0.5cm}
\centerline{
\psfrag{x11}{\raisebox{-0.0cm}{\hskip 0.0 true cm  (a)}}
\psfrag{x22}{\raisebox{-0.0cm}{\hskip 0.0 true cm  (b)}}
\psfrag{x33}{\raisebox{-0.0cm}{\hskip 0.0 true cm  (c)}}
\psfig{file=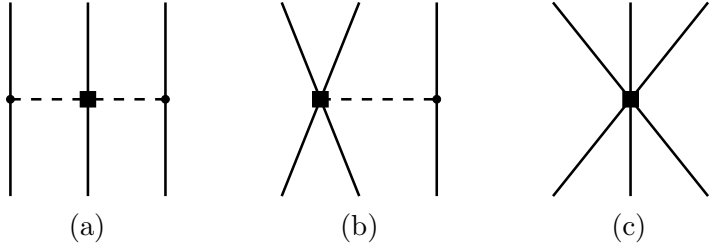,width=10cm}}
\vspace{0.3cm}
\centerline{\parbox{14cm}{
\caption[fig2]{\label{fig2} Subleading contribution to the 3NF at the order $(q/\Lambda )^3$. 
Solid rectangles refer to vertices with $\Delta_i = 1$.   For remaining notation see Fig.~\ref{fig1}.
}}}
\vspace{0.5cm}
\end{figure*}

The first nonvanishing 3NFs arise at order $(q/\Lambda )^3$ from the diagrams shown Fig.~\ref{fig2} with 
one subleading vertex of  dimension $\Delta_i = 1$.  
The contribution from the first graph in Fig.~\ref{fig2} is  
also incorporated in various phenomenological models like e.g. the TM99 3NF \cite{coon01} and 
given by  \cite{ko94} (see also \cite{FHK} for a related discussion):
\beq
\label{3nftpe}
V^{\rm 3N}_{2\pi}=\sum_{i \not= j \not= k} \frac{1}{2}\left(
  \frac{g_A}{2 F_\pi} \right)^2 \frac{( \vec \sigma_i \cdot \vec q_{i}) 
(\vec \sigma_j   \cdot \vec q_j  )}{(\vec q_i{}^2 + M_{\pi}^2 ) ( \vec
q_j{}^2 + M_{\pi}^2)}  F^{\alpha \beta}_{ijk} \tau_i^\alpha 
\tau_j^\beta \,,
\eeq
where  $\vec q_i \equiv \vec p_i \, ' - \vec p_i$; $\vec p_i$
($\vec p_i \, '$) are initial (final) momenta of the nucleon $i$ and 
\begin{displaymath}
F^{\alpha \beta}_{ijk} = \delta^{\alpha \beta} \left[ - \frac{4 \tilde c_1
    M_\pi^2}{F_\pi^2}  + \frac{2 c_3}{F_\pi^2}  
\vec q_i \cdot \vec q_j \right] + \sum_{\gamma} \frac{c_4}{F_\pi^2} \epsilon^{\alpha
\beta \gamma} \tau_k^\gamma  
\vec \sigma_k \cdot [ \vec q_i \times \vec q_j  ]\,.
\end{displaymath}
Here and in what follows, we use 
the usual notation to express the nuclear force: the quantity $V^{\rm 3N}_{2 \pi}$
is an operator with respect to spin and isospin quantum numbers and a matrix element with 
respect to momentum quantum numbers. Notice also that we have changed the notation of 
section \ref{sec:gen} and write the nucleon labels as subscripts of the spin
and isospin matrices (i.e. use $\fet \tau_i$ and $\vec \sigma_i$ instead of $\fet \tau (i)$ and 
$\vec \sigma (i)$), while the superscripts denote corresponding vector indices. 
Further, 
\beq
\tilde c_1 = c_1 + \frac{e^2 F^2_\pi f_1}{2 M_\pi^2}\,.
\eeq
Note that this renormalization of the sigma--term related LEC $c_1$ by the electromagnetic
LEC $f_1$  was already discussed in the analysis of pion--nucleon scattering \cite{fet01}.
Clearly, this electromagnetic shift of the LEC $c_1$ represents a higher--order effect 
and only needs to be taken into account at order $(q / \Lambda )^5$ and higher. 
The remaining contributions from graphs (b) and (c) in \ref{fig2} are given by \cite{E02}
\beqa
\label{3nfrest}
V^{\rm 3N}_{1\pi} &=& - \sum_{i \not= j \not= k} \frac{g_A}{8
  F_\pi^2} \, D \, \frac{\vec \sigma_i \cdot \vec q_i }{\vec q_i{}^2
  + M_\pi^2}  
\, \left( \fet \tau_j \cdot \fet \tau_i \right) 
(\vec \sigma_j \cdot \vec q_i ) \,, \nn
V^{\rm 3N}_{\rm cont} &=& \frac{1}{2} \sum_{j \not= k}  E \, ( \fet \tau_j \cdot \fet \tau_k ) \,,
\eeqa

\begin{figure*}[htb]
\vspace{0.5cm}
\centerline{
\psfrag{x33}{\raisebox{-0.0cm}{\hskip 0.0 true cm  (a)}}
\psfrag{x44}{\raisebox{-0.0cm}{\hskip 0.0 true cm  (b)}}
\psfrag{x55}{\raisebox{-0.0cm}{\hskip 0.0 true cm  (c)}}
\psfrag{x66}{\raisebox{-0.0cm}{\hskip 0.0 true cm  (d)}}
\psfig{file=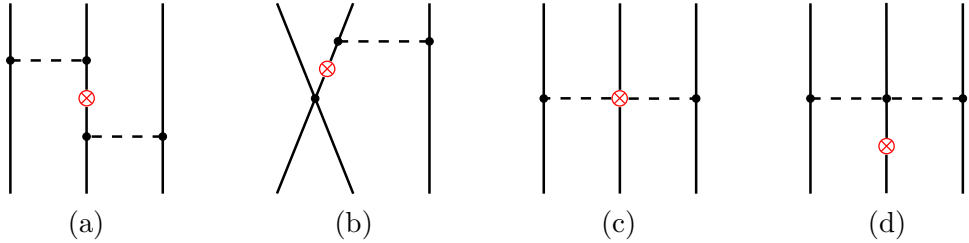,width=13.0cm}}
\vspace{0.3cm}
\centerline{\parbox{14cm}{
\caption[fig3]{\label{fig3} Leading isospin--violating contribution to the 3NF at the order $(q/\Lambda )^4$. 
Crossed circles refer to isospin--breaking vertices with $\Delta_i = 2$.   For remaining notation see Fig.~\ref{fig1}.
}}}
\vspace{0.5cm}
\end{figure*}

First isospin--conserving corrections to the 3NF arise at order $(q/\Lambda )^4$, where one has to consider 
tree diagrams with one vertex of the dimension $\Delta_i = 2$ as well various one--loop diagrams 
with the leading vertices. Derivation of these corrections to the 3NF will be published elsewhere. 
The main focus of the present work is related to isospin--breaking corrections which first appear at 
the same order $(q/\Lambda )^4$ and are given by the graphs in Fig.~\ref{fig3}. 
The first two diagrams (a) and (b) 
and the last one (d) 
are due to strong nucleon 
mass shift and of the order $\epsilon (q/\Lambda )^3 \sim (q/\Lambda )^4$. It should be understood that the proton--to--neutron 
mass difference has to be taken into account not only for intermediate but also for incoming and outgoing nucleon states.   
The corresponding corrections to the two--nucleon 
force have been recently studied in \cite{friar03,friar04}. In what follows, we will not separate the electromagnetic 
and strong shifts in the nucleon mass and express the result in terms of the proton--to--neutron mass difference
$\delta m = m_p - m_n$. We use the method of unitary transformation as detailed in \cite{EGM98} to calculate the 
relevant 3NF contributions. 
Utilizing the notation of this reference, the corresponding two--pion exchange potential can be written as:
\beqa
\label{operTPE}
V_{2 \pi} &=& \eta ' \bigg[ \frac{1}{2} H_{1} \frac{\lambda^1}{(H_0 - E_{\eta '})}  H_{1} \, \tilde \eta 
\,  H_{1} \frac{\lambda^1}{(H_0 - E_{\tilde \eta} )( H_0 - E_{\eta '} )}  H_{1} \nn 
&& \mbox{\hskip 0.7 true cm} -\frac{1}{8} H_{1} \frac{\lambda^1}{(H_0 - E_{\eta '})}  H_{1} \, \tilde \eta 
\,  H_{1} \frac{\lambda^1}{(H_0 - E_{\tilde \eta} )( H_0 - E_{\eta} )}  H_{1} \nn
&&  \mbox{\hskip 0.7 true cm} + \frac{1}{8} H_{1} \frac{\lambda^1}{(H_0 - E_{\eta '}) ( H_0 - E_{\tilde \eta} )}  
H_{1} \, \tilde \eta 
\,  H_{1} \frac{\lambda^1}{(H_0 - E_{\tilde \eta} )}  H_{1} \nn
&&  \mbox{\hskip 0.7 true cm} - 
\frac{1}{2} H_{1} \frac{\lambda^1}{(H_0 - E_{\eta})}  H_{1} \,  \frac{\lambda^2}{(H_0 - E_{\eta})} 
\,  H_{1} \frac{\lambda^1}{(H_0 - E_{\eta} )}  H_{1}
\bigg] \eta  + \mbox{h.~c.}
\eeqa
Here $\eta$, $\eta '$ and $\tilde \eta$ denote the projectors on the purely nucleonic subspace of the Fock space, while 
$\lambda^i$ refers to the projector on the states with $i$ pions. Further, $H_1$ is the leading 
$\pi NN$  vertex corresponding to the third term in the first line of eq.~(\ref{piN_2}), 
$H_0$ denotes the free Hamilton operator for pions and nucleons corresponding to the density
\beq
\label{freeH}
\mathcal{H}_0 = \frac{1}{2} \dot{\fet \pi} ^2 + 
\frac{1}{2} (\vec \nabla \fet \pi )^2 + \frac{1}{2} M_\pi^2  \fet \pi^2 +
\frac{1}{2} N^\dagger \delta m \tau_3 N\,,
\eeq
and $E_\eta$, $E_{\eta '}$ and $E_{\tilde \eta}$ refer to the energy of the nucleons in the states $\eta$, $\eta '$ and 
$\tilde \eta$, respectively.   
Notice that the first three terms in eq.~(\ref{operTPE}) subsume the contributions of the reducible graphs while 
the last term refers to the irreducible topology. Neglecting the proton--to--neutron mass difference in eq.~(\ref{freeH})
one recovers the isospin symmetric result of \cite{EGM98}:
\beq
V_{2 \pi} = \eta ' \bigg[ \frac{1}{2} H_{1} \frac{\lambda^1}{(\omega)^2}  H_{1} \, \tilde \eta \, H_{1} 
\frac{\lambda^1}{\omega}  H_{1}  
+ \frac{1}{2} H_{1} \frac{\lambda^1}{\omega}  H_{1} \, \tilde \eta \, H_{1} \frac{\lambda^1}{(\omega )^2}  H_{1} - 
 H_{1} \frac{\lambda^1}{\omega}  H_{1} \frac{\lambda^2}{\omega_1 + \omega_2} H_{1} \frac{\lambda^1}{(\omega )^2}  H_{1} 
\bigg] \eta \,,
\eeq
where $\omega$ denotes the pionic free energy. We remark that eq.~(\ref{operTPE}) can also be used to 
calculate relativistic $1/m$--corrections to the two--pion exchange potential 
if one keeps the nucleon kinetic energy term in eq.~(\ref{freeH}). An additional unitary transformation should,
however, be performed in order to end up with the potential used in \cite{N3LO}.

Explicit evaluation of the 3NF using eq.~(\ref{operTPE}) leads to the following result:
\beqa
\label{3NFisosp1}
V^{\rm 3N}_{2\pi} &=& \sum_{i \not= j \not= k} \,2 \delta m  \,  \left(
  \frac{g_A}{2 F_\pi} \right)^4 \frac{( \vec \sigma_i \cdot \vec q_{i}  ) 
(\vec \sigma_j \cdot \vec q_j  )}{(\vec q_i{} ^2 + M_{\pi}^2 )^2 ( \vec
q_j{} ^2 + M_{\pi}^2)} \bigg\{ [\vec q_i \times \vec q_j ] \cdot \vec \sigma_k  \, [ \fet \tau_i \times \fet \tau_j ]^3
 \nn
&& {} + \vec q_i \cdot \vec q_j \left[ (\fet \tau_i \cdot \fet \tau_k ) \tau_j^3  -
(\fet \tau_i \cdot \fet \tau_j ) \tau_k^3 \right] \bigg\}\,.
\eeqa
Notice that we have expanded the energy denominators in powers of $\delta m$ in eq.~(\ref{operTPE}) and kept only 
the linear terms. Similarly to the case of the two--nucleon potential \cite{friar03}, the resulting 3NF is entirely due to 
irreducible diagrams.  
As a cross--check of our approach, we have also calculated the two--pion exchange 2NF corresponding 
to eq.~(\ref{operTPE}) and recovered the results of \cite{friar03}. 

The contribution of the one--pion exchange diagram (b) in Fig.~\ref{fig3} is given by the operators
\beq
\label{operOPE}
V_{1 \pi} = \eta ' \bigg[ - \frac{1}{2} H_{1} \frac{\lambda^1}{(H_0 - E_{\eta}) (H_0 - E_{\tilde \eta})}  H_{1} \, \tilde \eta \,
H_2 + \frac{1}{2} H_{1} \frac{\lambda^1}{(H_0 - E_{\eta}) }  H_{2}   \frac{\lambda^1}{(H_0 - E_{\eta}) } H_1 
\bigg] \eta  + \mbox{h.~c.}
\eeq
where $H_2$ corresponds to the first two terms in eq.~(\ref{NN2}). Similarly to the previously considered case,
we recover the result of \cite{EGM98} in the limit $\delta m \rightarrow 0$:
\beq
V_{1 \pi} = \eta ' \bigg[ - \frac{1}{2} H_{1} \frac{\lambda^1}{( \omega )^2 }  H_{1} \, \tilde \eta \, H_2 
- \frac{1}{2} H_{2}\,  \tilde \eta \, H_1 \frac{\lambda^1}{( \omega )^2 }  H_{1}  +
H_{1} \frac{\lambda^1}{\omega }  H_{2}   \frac{\lambda^1}{\omega} H_1 \bigg] \eta
\eeq
We find the following expression for the isospin--breaking one--pion exchange 3NF
\beq
\label{3NFisosp2}
V^{\rm 3N}_{1\pi} = \sum_{i \not= j \not= k} 2 \, \delta m \, C_T   \left( \frac{g_A}{2 F_\pi} \right)^2
\frac{\vec \sigma_i \cdot \vec q_i}{(\vec q_i {} ^2 + M_{\pi}^2)^2} \, [ \fet \tau_k \times \fet \tau_i ]^3 \; 
[ \vec \sigma_j \times \vec \sigma_k ] \cdot \vec q_i \,.
\eeq
Notice that $V^{\rm 3N}_{1\pi}$ can be rewritten in an equivalent form making use of the relation
\beq
[ \fet \tau_k \times \fet \tau_i ]^3 \; 
[ \vec \sigma_j \times \vec \sigma_k ] \cdot \vec q_i = 
\left( ( \fet \tau_i \cdot \fet \tau_j ) \tau_k^3 - (\fet \tau_i \cdot \fet \tau_k ) \tau_j^3 \right) 
(\vec  \sigma_j \cdot \vec q_i )\,,
\eeq
which holds true when the corresponding operators act on antisymmetrized states with respect to  $j$ and $k$. 

The diagram (c) in Fig.~\ref{fig3} is due to the $c_5$--term in eq.~(\ref{piN_2}) and of the order 
$\epsilon (q/\Lambda )^3 \sim (q/\Lambda )^4$ as well. Denoting the interaction $\propto c_5$ by $H_3$,
the contribution of this graph is given by 
\beq
\label{oper_c5}
V_{2 \pi} = \eta ' \bigg[ H_{1} \frac{\lambda^1}{\omega}  H_{3} \frac{\lambda^1}{\omega }  H_{1}  
+  H_{1} \frac{\lambda^1}{\omega}  H_{1} \frac{\lambda^2}{(\omega_1 + \omega_2 )}  H_{3} 
+    H_{3} \frac{\lambda^2}{(\omega_1 + \omega_2 )}  H_{1} \frac{\lambda^1}{\omega} H_{1} \bigg]
\eta \,.
\eeq
Alternatively, one can use the Feynman graph technique to evaluate the corresponding 3NF. We find
\beq
\label{3NFisosp3}
V_{2 \pi}^{\rm 3N} = \sum_{i \not= j \not= k} \, \frac{ ( \delta m )^{\rm str.}}{4 F_\pi^2}  \left(
  \frac{g_A}{2 F_\pi} \right)^2 \, \frac{( \vec \sigma_i \cdot \vec q_{i} ) 
(\vec \sigma_j   \cdot \vec q_j  )}{(\vec q_i{}^2 + M_{\pi}^2 ) ( \vec
q_j{}^2 + M_{\pi}^2)} (\fet \tau_i \cdot \fet \tau_k ) \tau_j^3\,.
\eeq
Notice that all leading (i.e.~$\sim (q/\Lambda )^4$) isospin--violating 3NFs given by 
eqs.~(\ref{3NFisosp1}), (\ref{3NFisosp2}) and (\ref{3NFisosp3}) are charge--symmetry--breaking,
i.e.~ of class (III) in the notation of section  \ref{sec:3N}. We further point out that 
although the $M_{\pi^\pm} \neq M_{\pi^0}$--corrections to the graphs in Fig.~\ref{fig1} 
given by the first three graphs (a), (b) and (c) in the later 
Fig.~\ref{fig6} are formally also of the order 
$(q/\Lambda )^4$, they  lead to $1/m$--suppressed contributions to the 3NF for the same reason 
as do the corresponding isospin--conserving terms. 

The contribution of the last diagram (d) in Fig.~\ref{fig3} is given by 
\beqa
\label{operTPEmissed}
V_{2 \pi} &=& \frac{1}{2} \eta ' \bigg[ H_{1} \frac{\lambda^1}{(H_0 - E_{\eta})}  H_{1} 
\frac{\lambda^2}{(H_0 - E_{\eta})}
H_1^{\rm WT} + H_1^{\rm WT} \frac{\lambda^2}{(H_0 - E_{\eta})} H_{1} \frac{\lambda^1}{(H_0 - E_{\eta})}  H_{1} \nn
&&{} \mbox{\hskip 0.8 true cm} + H_{1} \frac{\lambda^1}{(H_0 - E_{\eta}) }  H_{1}^{\rm WT}   
\frac{\lambda^1}{(H_0 - E_{\eta}) } H_1 \bigg] \eta  + \mbox{h.~c.}\,,
\eeqa
where $H_1^{\rm WT}$ refers to the Weinberg--Tomozawa vertex. Explicit evaluation of this graph can be performed
expanding the above expression in powers of $\delta m$ and keeping the terms $\propto \delta m$. Alternatively,
one can use the Feynman graph technique. In that case one should use for the energy transfer of the nucleon $i$:
$q_i^0 = (p_i ' )^0 - p_i^0 = \Delta m + \mathcal{O} (m^{-1})$, where $\Delta m$ denotes the nucleon mass difference 
in the final and initial state. We find:
\beq
\label{3NFmissing}
V_{2 \pi}^{\rm 3N} = \sum_{i \not= j \not= k} \, \frac{ \delta m }{4 F_\pi^2}  \left(
  \frac{g_A}{2 F_\pi} \right)^2 \, \frac{( \vec \sigma_i \cdot \vec q_{i} ) 
(\vec \sigma_j   \cdot \vec q_j  )}{(\vec q_i{}^2 + M_{\pi}^2 ) ( \vec
q_j{}^2 + M_{\pi}^2)} \left[ (\fet \tau_i \cdot \fet \tau_k ) \tau_j^3 - (\fet \tau_i \cdot \fet \tau_j ) \tau_k^3 \right]\,.
\eeq

\begin{figure*}[htb]
\vspace{0.5cm}
\centerline{
\psfrag{x11}{\raisebox{-0.0cm}{\hskip 0.0 true cm  (a)}}
\psfrag{x22}{\raisebox{-0.0cm}{\hskip 0.0 true cm  (b)}}
\psfrag{x33}{\raisebox{-0.0cm}{\hskip 0.0 true cm  (c)}}
\psfrag{x44}{\raisebox{-0.0cm}{\hskip 0.0 true cm  (d)}}
\psfrag{x55}{\raisebox{-0.0cm}{\hskip 0.0 true cm  (e)}}
\psfrag{x66}{\raisebox{-0.0cm}{\hskip 0.0 true cm  (f)}}
\psfig{file=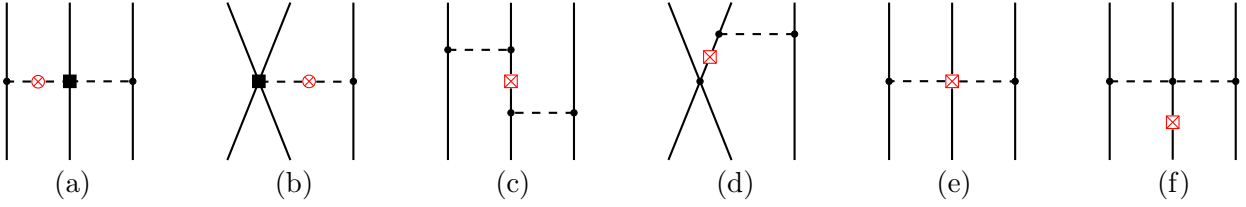,width=16.7cm}}
\vspace{0.3cm}
\centerline{\parbox{14cm}{
\caption[fig4]{\label{fig4} Subleading isospin--violating contribution to the 3NF at the order $(q/\Lambda )^5$. 
Crossed rectangles refer to isospin--breaking vertices with $\Delta_i = 3$.   For remaining notation see Fig.~\ref{fig1}.
}}}
\vspace{0.5cm}
\end{figure*}

The first corrections to the leading isospin--breaking 3NFs arise  from the diagrams 
(a) (b) and (e) in Fig.~\ref{fig4} and are of  the order $(e/4 \pi)^2 q/\Lambda \sim (q/\Lambda )^5$ .
Notice that the contributions of the graphs 
(c), (d) and (f) in this figure 
are already included in eqs.~(\ref{3NFisosp1}), 
(\ref{3NFisosp2}) and (\ref{3NFmissing}). 
The first two graphs in  Fig.~\ref{fig4} represent isospin--violating corrections to the graphs (a) and (b) in 
Fig.~\ref{fig2} due to the pion mass difference and lead to
\beqa
\label{3NFisosp4}
V^{\rm 3N}_{2\pi}&=&\sum_{i \not= j \not= k} \, \delta M_\pi^2 \, \left(
  \frac{g_A}{2 F_\pi} \right)^2 \frac{( \vec \sigma_i \cdot \vec q_{i})
(\vec \sigma_j   \cdot \vec q_j  )}{(\vec q_i{}^2 + M_{\pi}^2 )^2 ( \vec
q_j{}^2 + M_{\pi}^2)}  \bigg\{ \tau_i^3 \tau_j^3 \left[  - 
\frac{4 c_1 M_\pi^2}{F_\pi^2} +    \frac{2 c_3}{F_\pi^2} (\vec q_i \cdot \vec q_j )   \right] \nn
&& \mbox{\hskip 1.7 true cm} 
+ \frac{c_4}{F_\pi^2} \, \tau_i^3 \, 
[\fet \tau_j \times \fet \tau_k ]^3 \, [\vec q_i \times \vec q_j ] \cdot \vec \sigma_k \bigg\} \nn
V^{\rm 3N}_{1\pi}&=&- \sum_{i \not= j \not= k} \, \delta M_\pi^2 \, \frac{g_A}{8 F_\pi^2} \, D \, 
\frac{\vec \sigma_i \cdot \vec q_i }{(\vec q_i{}^2  + M_\pi^2 )^2}  
\,  \tau_i^3 \tau_j^3 (\vec \sigma_j \cdot \vec q_i )\,,
\eeqa
where we have defined 
\beq
\delta M_\pi^2 =  M_{\pi^\pm}^2 - M_{\pi^0}^2\,.
\eeq
Notice that at this order (i.e. at $(q/\Lambda )^5$) one has to distinguish between 
the charged and neutral pion masses in the pion propagators in eqs.~(\ref{3nftpe}) and (\ref{3nfrest}). 
Isospin--violating corrections in eq.~(\ref{3NFisosp4}) are consistent with taking $M_{\pi^\pm}$ in the 
pion propagators in eqs. ~(\ref{3nftpe}) and (\ref{3nfrest}).
The contribution of the diagram (e) can be obtained from eq.~(\ref{oper_c5}):
\beqa
\label{3NFisosp5}
V^{\rm 3N}_{2\pi} &=&\sum_{i \not= j \not= k} \, \left(
  \frac{g_A}{2 F_\pi} \right)^2 \frac{( \vec \sigma_i \cdot \vec q_{i})
(\vec \sigma_j   \cdot \vec q_j  )}{(\vec q_i{}^2 + M_{\pi}^2 ) ( \vec
q_j{}^2 + M_{\pi}^2)}  \bigg\{  \frac{(\delta m )^{\rm em.}}{4 F_\pi^2} 
\left( (\fet \tau_i \cdot \fet \tau_j ) \tau_k^3  - (\fet \tau_i \cdot \fet \tau_k ) \tau_j^3 \right)  \nn 
&& \mbox{\hskip 1.7 true cm} 
+ f_1 e^2 \, \tau_i^3 \tau_j^3 
\bigg\}\,.
\eeqa
The 3NFs resulting from $M_{\pi^\pm} \neq M_{\pi^0}$ in graphs (a) and (b) of Fig.~\ref{fig4} are 
charge--symmetry--conserving 
(i.e.~ class (II)) while the diagram (e) in this figure gives rise to both charge--symmetry--conserving 
($\propto f_1$) and charge--symmetry--breaking ($\propto (\delta m )^{\rm em.}$) 3NFs. 
We stress again that the contribution $\sim f_1$ is considered here for the first time.
Notice further that the charge--symmetry breaking 3NFs in eqs.~(\ref{3NFisosp3}), (\ref{3NFmissing}) 
and (\ref{3NFisosp5}) can be combined to:
\beq
\label{3NFprom}
V_{2 \pi}^{\rm 3N} = \sum_{i \not= j \not= k} \, \frac{( \delta m )^{\rm str.}}{4 F_\pi^2}  \left(
  \frac{g_A}{2 F_\pi} \right)^2 \, \frac{( \vec \sigma_i \cdot \vec q_{i} ) 
(\vec \sigma_j   \cdot \vec q_j  )}{(\vec q_i{}^2 + M_{\pi}^2 ) ( \vec
q_j{}^2 + M_{\pi}^2)} \left[ 2 (\fet \tau_i \cdot \fet \tau_k ) \tau_j^3 - (\fet \tau_i \cdot \fet \tau_j ) \tau_k^3 \right]\,.
\eeq
The coordinate space representation of the obtained 3NFs is given in appendix \ref{sec:coord}.
Notice that there exist further diagrams at this order which, however, lead to vanishing contributions 
and are not considered in the present work.

Let us now comment on the obtained results. First of all, we notice a (formally) larger relative size 
of the isospin--breaking corrections compared to the two--nucleon sector. 
Indeed, isospin--breaking 3NFs are suppressed by  
$q/\Lambda$ compared to the isospin--conserving 3NFs, while the suppression factor in case of 
the 2NF is  $( q/\Lambda )^2$. Secondly, the 
leading isospin--breaking corrections to the 2N and 3N forces arise from different sources. 
In particular, the dominant contribution to the 3NF is governed by the proton--to--neutron mass difference,
which only gives a sub--subleading isospin--breaking correction to the 2N force. Further,
charge dependence of the pion--nucleon coupling constant does not show up in the 3NF at the 
considered order. Similarly, the leading isospin--breaking 3N contact interaction is of the order 
$\epsilon M_\pi^2 ( q/\Lambda )^3 \sim  ( q/\Lambda )^6$ and therefore does not need to be included. 
Last but not least, we notice that the hierarchy of isospin--violating forces observed in the two--nucleon 
system (i.e. charge--independence--breaking forces are stronger than charge--symmetry--breaking forces \cite{kolck}) 
is not valid for three--nucleon forces.

\subsection{Estimation of the size of the isospin--breaking 3NFs}
\label{sec:est}

\begin{figure*}[htb]
\vspace{0.5cm}
\centerline{
\psfrag{x11}{\raisebox{-0.0cm}{\hskip 0.0 true cm  (a)}}
\psfrag{x22}{\raisebox{-0.0cm}{\hskip 0.0 true cm  (b)}}
\psfrag{x33}{\raisebox{-0.0cm}{\hskip 0.0 true cm  (c)}}
\psfrag{x44}{\raisebox{-0.0cm}{\hskip 0.0 true cm  (d)}}
\psfig{file=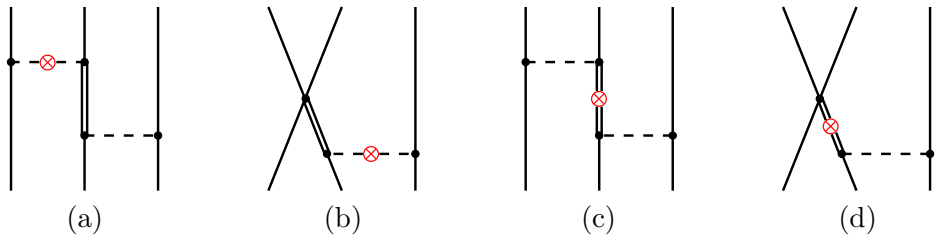,width=13.0cm}}
\vspace{0.3cm}
\centerline{\parbox{14cm}{
\caption[fig5]{\label{fig5} Isospin--violating contribution to the 3NF due to intermediate 
$\Delta$--excitation (double lines), which are not considered explicitly in the present work.   
The effect of such diagrams is hidden in certain contact operators that originates from integrating
out the delta in the approach considered here.
For remaining notation see Fig.~\ref{fig1}.  
}}}
\vspace{0.5cm}
\end{figure*}

Having derived the dominant isospin--breaking 3NF corrections it would be very interesting to see how large 
the effects actually are. This, however, requires explicit calculations of few--nucleon observables,
which goes beyond the scope of the present study. Here we restrict ourselves to the following very rough 
estimation. Consider the two--pion--exchange correction given in eq.~(\ref{3NFisosp1}). 
Approximating $1/(\vec q_i {}^2 + M_\pi^2) \sim 1/M_\pi^2$ we obtain the same spin--space structures as 
the ones which enter the leading isospin conserving 3NF in eq.~(\ref{3nftpe}). Neglecting the isospin structure 
one observes that the strength of the isospin--breaking terms in 
eqs.~(\ref{3NFisosp1}), (\ref{3NFisosp3}) and 
(\ref{3NFmissing})
reaches few percent 
of the strength of the corresponding isospin--conserving pieces in  eq.~(\ref{3nftpe}). 
Based on the above estimates and on the fact that two--pion exchange 3NFs typically contribute several 
hundreds keV to the binding 
energy of $^3$H  and $^3$He,\footnote{In \cite{kam01} contributions of various pieces of the 
Tucson--Melbourne 3NF to the $^3$H are considered. While the so--called $a$--term 
(it corresponds to the $c_1$--term in the chiral 3NF) was found to provide only a tiny contribution, 
the $b$-- ($\propto c_3$) and $d$--terms ($\propto c_4$) give about $250 \ldots 300$ keV each. 
In the analysis \cite{E02} based on chiral EFT, the expectation value of the two--pion exchange 3NF 
for $^3$H (with the reduced values of the LECs $c_{3,4}$) 
was found to be $390 \ldots 730$ keV depending on the cut--off chosen.}
one might expect the contribution of the isospin--breaking 3NF in eq.~(\ref{3NFisosp1}) to 
the $^3$He--$^3$H binding--energy difference to reach $10 \ldots 20$ keV. 
On the other hand, 
the relative strength of the formally subleading two--pion exchange terms in eqs.~(\ref{3NFisosp4}) and (\ref{3NFisosp5}) 
reaches even $2 \delta M_\pi^2 / M_\pi^2 \sim 15\%$. This surprisingly large size of the subleading  
isospin--breaking corrections compared to the leading ones is due to  
the LECs $c_{1,3,4}$, which enter eq.~(\ref{3NFisosp4}) and are numerically 
large (the physics behind this enhancement of the LECs is well understood \cite{BKMlec}\footnote{In that paper is was
shown, that the smallness of the $N\Delta$ mass splitting enhances certain pion-nucleon LECs when one
integrates out the delta. Furthermore, scalar and vector mesons make large contributions to $c_1$ and $c_4$,
respectively.}). 
Notice that a similar situation occurs for the isospin--conserving 
two--pion exchange 2N force, where the numerically dominant contributions are provided by subleading terms. 
One should, however, keep in mind that the isospin--breaking 3NFs $\propto c_{1,3,4}$ do not lead to 
charge--symmetry--breaking and thus do not contribute i.e.~to the $^3$He--$^3$H binding--energy difference.
The leading charge--symmetry--breaking one--pion--exchange 3NF in eq.~(\ref{3NFisosp2}) 
is numerically smaller in size than the corresponding subleading charge--symmetry--conserving contribution 
in eq.~(\ref{3NFisosp4}) as well, although the reason is now completely different. 
The 3NF in eq.~(\ref{3NFisosp2}) is proportional 
to the LEC $C_T$ which is numerically small \cite{EMGE}.\footnote{In EFT without or with perturbative pions, 
one has $C_T=0$ in the limit when both NN S--wave scattering lengths go to infinity \cite{meh99}.}
It should be noted in this context that the size of the isospin--breaking 3NFs would be more natural if one 
would treat $\Delta$--isobar as an explicit degree of freedom. In that case a 
large portion of the subleading 3NFs $\propto c_{3, 4}$ and $D$ due to 
graphs (a) and (b) in Fig.~\ref{fig5} would be promoted to the leading order.
Note also that such an approach with explicit deltas is much more complicated
since one has to deal with more structures and also needs e.g. to reanalyze
pion-nucleon scattering (for an attempt see e.g. \cite{fetdelta}).
Further, one should keep in mind that the above numerical estimates are very rough. In particular, 
taking into account the neglected isospin structure will change the numbers by several times 
depending on the process considered. Thus, only explicit calculation of various few--nucleon observables 
will provide quantitative insights on the size of the derived 3NFs.

\begin{figure*}[htb]
\vspace{0.5cm}
\centerline{
\psfrag{x11}{\raisebox{-0.0cm}{\hskip 0.0 true cm  (a)}}
\psfrag{x22}{\raisebox{-0.0cm}{\hskip 0.0 true cm  (b)}}
\psfrag{x33}{\raisebox{-0.0cm}{\hskip 0.0 true cm  (c)}}
\psfrag{x44}{\raisebox{-0.0cm}{\hskip 0.0 true cm  (d)}}
\psfrag{x55}{\raisebox{-0.0cm}{\hskip 0.0 true cm  (e)}}
\psfig{file=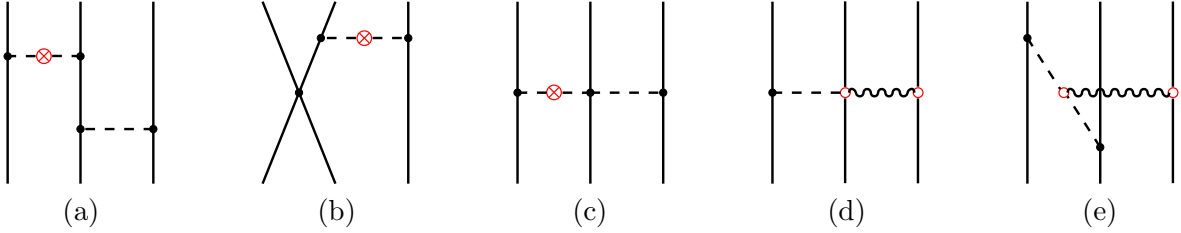,width=16.5cm}}
\vspace{0.3cm}
\centerline{\parbox{14cm}{
\caption[fig6]{\label{fig6} Selected $1/m$--corrections to the isospin--violating 3NF, which are 
not considered in the present work. Wavy lines refer to photons and unfilled circles denote vertices with 
photons.     For remaining notation see Fig.~\ref{fig1}.
}}}
\vspace{0.5cm}
\end{figure*}

Finally, we point out that there are many $1/m$--corrections to the obtained results,
some of which  are depicted in Fig.~\ref{fig6}. Since we consider the 
nucleon mass as a larger scale compared to $\Lambda$, such relativistic corrections are irrelevant at the 
order considered in this work. Notice, however, that if one would adopt the counting rule $m \sim \Lambda$, various 
$1/m$--corrections (including the ones due to virtual photons) would have to be included at the subleading 
order $(q/\Lambda)^5$. Some 3NF diagrams due to virtual photon exchange have been considered 
by Yang and found to provide relatively small contributions 
of the order of $\sim 7$ keV to the $^3$He--$^3$H binding--energy difference \cite{yang79,yang83}.
Furthermore, we remind the reader that the long--range electromagnetic 
3NFs might, in principle, give rise to large contributions to scattering observables 
under certain kinematic conditions \cite{N3LO}.

\section{Summary}
\def\theequation{\arabic{section}.\arabic{equation}}
\label{sec:summary}

Here, we summarize the pertinent results of this investigation.

\begin{itemize}
\item[i)] We have given a classification scheme for $A$--nucleon forces according
to their isospin dependence. In the 3N system, one finds three different classes of forces,
according to their transformation properties under isospin and charge--symmetry 
transformations.
\item[(ii)] We have worked out the leading and subleading isospin--violating 3NFs.
The leading contributions are generated by one-- and two--pion exchange diagrams
with their strength given by the strong neutron--proton mass difference. The
subleading corrections are again given by one-- and two--pion exchange diagrams,
driven largely by the charged--to--neutral pion mass difference and also
by the electromagnetic neutron--proton mass difference and the dimension two
electromagnetic LEC $f_1$, that plays an important role in the pion--nucleon system. 
\item[(iii)] We have estimated the relative strength of the leading and subleading
corrections compared to the isospin--conserving 3NF at the same order. Isospin--violating 
3NFs are expected to provide a small but non--negligible contribution to the 
$^3$He--$^3$H binding--energy difference.
\end{itemize}

In the future, these isospin--breaking forces should be used to analyze three- and
four--nucleon systems based on chiral EFT, extending e.g. the work presented in
\cite{E02}. 

\section*{Acknowledgments}

We are grateful to Jerry Miller for useful discussions and to Walter Gl\"ockle for 
his careful reading and comments on this paper. We are also grateful to Bira van 
Kolck for some constructive criticism on the first version of the manuscript.
This work has been supported by the 
U.S.~Department of Energy Contract No.~DE-AC05-84ER40150 under which the 
Southeastern Universities Research Association (SURA) operates the Thomas Jefferson 
National Accelerator Facility and by the Deutsche Forschungsgemeinschaft
through funds provided to the SFB/TR 16 ``Subnuclear Structure of Matter''.

\bigskip

\appendix
\def\theequation{\Alph{section}.\arabic{equation}}
\setcounter{equation}{0}
\section{Coordinate space representation}
\label{sec:coord}

The leading and subleading 3NFs 
are local and can easily be transformed into 
coordinate space. We first define the following operators:
\beqa
\label{o1}
O^1_{ijk} &=& \int \frac{d^3 q_i}{(2 \pi )^3}  \frac{d^3 q_j}{(2 \pi )^3} \, e^{ i \vec q_i \cdot \vec r_{ik}}\, 
 e^{ i \vec q_j \cdot \vec r_{jk}} \, \frac{( \vec \sigma_i \cdot \vec q_{i}  ) 
(\vec \sigma_j \cdot \vec q_j  )}{(\vec q_i{} ^2 + M_{\pi}^2 )^2 ( \vec
q_j{} ^2 + M_{\pi}^2)} \, [\vec q_i \times \vec q_j ] \cdot \vec \sigma_k  \nn
&=& ( \vec \sigma_i \cdot \vec \nabla_{ik} ) ( \vec \sigma_j \cdot \vec \nabla_{jk} ) 
[ \vec \nabla_{ik} \times \vec \nabla_{jk} ] \cdot \vec \sigma_k \, h_2 (r_{ik} ) \, h_1 ( r_{jk} )\nn
&=& \frac{1}{32 \pi^2}\,  \frac{e^{-x_{ik}}}{r_{ik}} \,\frac{e^{-x_{jk}}}{r_{jk}^3}  
\Big( ( \vec \sigma_i \cdot \vec {\hat r}_{ik} )  ( \vec \sigma_j \cdot \vec { \hat r}_{jk} )\, 
[\vec {\hat r}_{ik} \times \vec {\hat r} _{jk} ] \cdot \vec \sigma_k  \, (1 + x_{ik} ) ( 3 + 3 x_{jk} + x_{jk}^2 ) \nn
&& {} \mbox{\hskip 3.2 true cm} 
+ [ \vec \sigma_i \times \vec \sigma_j  ] \cdot \vec \sigma_k  \; ( 1 +  x_{jk})
 - ( \vec \sigma_i \cdot \vec {\hat r}_{ik} )  \; 
[\vec {\hat r}_{ik} \times \vec \sigma_j ] \cdot \vec \sigma_k  \; (1 + x_{ik} ) ( 1 +  x_{jk})  \nn
&& {} \mbox{\hskip 3.2 true cm} 
- [ \vec \sigma_i \times \vec {\hat r}_{jk} ] \cdot \vec \sigma_k \; ( \vec \sigma_j \cdot \vec {\hat r}_{jk} )  \; 
( 3 + 3 x_{jk} + x_{jk}^2 )  \Big) \nn
&& + \frac{1}{24 \pi}\,  \frac{e^{-x_{ik}}}{r_{ik}} \,\delta^3 ( r_{jk} ) \;
\Big(  ( \vec \sigma_i \cdot \vec {\hat r}_{ik} )  \; 
[\vec {\hat r}_{ik} \times \vec \sigma_j ] \cdot \vec \sigma_k  \; (1 + x_{ik} ) - 
[ \vec \sigma_i \times \vec \sigma_j  ] \cdot \vec \sigma_k  \Big) \\
O^2_{ijk} &=& \int \frac{d^3 q_i}{(2 \pi )^3}  \frac{d^3 q_j}{(2 \pi )^3} \, e^{ i \vec q_i \cdot \vec r_{ik}}\, 
 e^{ i \vec q_j \cdot \vec r_{jk}} \, \frac{( \vec \sigma_i \cdot \vec q_{i}  ) 
(\vec \sigma_j \cdot \vec q_j  )}{(\vec q_i{} ^2 + M_{\pi}^2 )^2 ( \vec
q_j{} ^2 + M_{\pi}^2)} \, ( \vec q_i \cdot  \vec q_j )  \nn
&=& ( \vec \sigma_i \cdot \vec \nabla_{ik} ) ( \vec \sigma_j \cdot \vec \nabla_{jk} ) 
( \vec \nabla_{ik} \cdot \vec \nabla_{jk} ) \, h_2 (r_{ik} ) \, h_1 ( r_{jk} )\nn
&=& \frac{1}{32 \pi^2}\,  \frac{e^{-x_{ik}}}{r_{ik}} \,\frac{e^{-x_{jk}}}{r_{jk}^3}  
\Big( ( \vec \sigma_i \cdot \vec {\hat r}_{ik} )  ( \vec \sigma_j \cdot \vec { \hat r}_{jk} )\, 
(\vec {\hat r}_{ik} \cdot \vec {\hat r} _{jk} ) \, (1 + x_{ik} ) ( 3 + 3 x_{jk} + x_{jk}^2 ) \nn
&& {} \mbox{\hskip 3.2 true cm} 
+ ( \vec \sigma_i \cdot \vec \sigma_j  )  \; ( 1 +  x_{jk})
 - ( \vec \sigma_i \cdot \vec {\hat r}_{ik} )  \; 
( \vec \sigma_j \cdot \vec {\hat r}_{ik})  \; (1 + x_{ik} ) ( 1 +  x_{jk})  \nn
&& {} \mbox{\hskip 3.2 true cm} 
- ( \vec \sigma_i \cdot \vec {\hat r}_{jk} )\; ( \vec \sigma_j \cdot \vec {\hat r}_{jk} )  \; 
( 3 + 3 x_{jk} + x_{jk}^2 )  \Big) \nn
&& + \frac{1}{24 \pi}\,  \frac{e^{-x_{ik}}}{r_{ik}} \,\delta^3 ( r_{jk} ) \;
\Big(  ( \vec \sigma_i \cdot \vec {\hat r}_{ik} )  \; 
(\vec \sigma_j  \cdot \vec {\hat r}_{ik}) \; (1 + x_{ik} ) - 
(\vec \sigma_i \cdot \vec \sigma_j ) \Big) \\
O^3_{ijk} &=& \int \frac{d^3 q_i}{(2 \pi )^3}  \frac{d^3 q_j}{(2 \pi )^3} \, e^{ i \vec q_i \cdot \vec r_{ik}}\, 
 e^{ i \vec q_j \cdot \vec r_{jk}} \, \frac{( \vec \sigma_i \cdot \vec q_{i}  ) 
(\vec \sigma_j \cdot \vec q_j  )}{(\vec q_i{} ^2 + M_{\pi}^2 ) ( \vec
q_j{} ^2 + M_{\pi}^2)}   \nn
&=& - ( \vec \sigma_i \cdot \vec \nabla_{ik} ) ( \vec \sigma_j \cdot \vec \nabla_{jk} ) 
\, h_1 (r_{ik} ) \, h_1 ( r_{jk} )\nn
&=& \frac{1}{16 \pi^2}\,  \frac{e^{-x_{ik}}}{r_{ik}^2} \,\frac{e^{-x_{jk}}}{r_{jk}^2}  
( \vec \sigma_i \cdot \vec {\hat r}_{ik} )  ( \vec \sigma_j \cdot \vec { \hat r}_{jk} )\,
 (1 + x_{ik} )  (1 + x_{jk} ) \\
O^4_{ijk} &=& \int \frac{d^3 q_i}{(2 \pi )^3}  \frac{d^3 q_j}{(2 \pi )^3} \, e^{ i \vec q_i \cdot \vec r_{ik}}\, 
 e^{ i \vec q_j \cdot \vec r_{jk}} \, \frac{( \vec \sigma_i \cdot \vec q_{i}  ) 
(\vec \sigma_j \cdot \vec q_j  )}{(\vec q_i{} ^2 + M_{\pi}^2 )^2 ( \vec
q_j{} ^2 + M_{\pi}^2)}   \nn
&=& - ( \vec \sigma_i \cdot \vec \nabla_{ik} ) ( \vec \sigma_j \cdot \vec \nabla_{jk} ) 
\, h_2 (r_{ik} ) \, h_1 ( r_{jk} )\nn
&=& \frac{1}{32 \pi^2}\,  e^{-x_{ik}} \,\frac{e^{-x_{jk}}}{r_{jk}^2}  
( \vec \sigma_i \cdot \vec {\hat r}_{ik} )  ( \vec \sigma_j \cdot \vec { \hat r}_{jk} )\,
 (1 + x_{jk} ) \\
\label{o5}
O^5_{ijk} &=& \int \frac{d^3 q_i}{(2 \pi )^3}  \frac{d^3 q_j}{(2 \pi )^3} \, e^{ i \vec q_i \cdot \vec r_{ik}}\, 
 e^{ i \vec q_j \cdot \vec r_{jk}} \, \frac{( \vec \sigma_i \cdot \vec q_{i} )  
}{(\vec q_i{} ^2 + M_{\pi}^2 )^2}  \, ( \vec \sigma_j  \cdot \vec q_i )  \nn
&=& - ( \vec \sigma_i \cdot \vec \nabla_{ik} )  ( \vec \sigma_j  \cdot \vec \nabla_{ik} )
\, h_2 (r_{ik} ) \, g ( r_{jk} )\nn
&=& - \frac{1}{8 \pi}\,  \frac{e^{-x_{ik}}}{r_{ik}} \, \delta^3 ( r_{jk} )   \;
\bigg( ( \vec \sigma_i \cdot \vec {\hat r}_{ik} )  ( \vec \sigma_j \cdot \vec { \hat r}_{ik} )\,
 (1 + x_{jk} )  - ( \vec \sigma_i \cdot \vec \sigma_j ) \bigg)\,.
\eeqa
Here $\vec r_{ij}$ is the relative distance between the nucleons $i$ and $j$,
$r_{ij} = | \vec r_{ij} |$, $\vec {\hat r}_{ij} = \vec r_{ij}/r_{ij}$ and $x_{ij} = M_\pi r_{ij}$. 
Further, 
\beqa
h_1 (r ) &=& \int \frac{d^3 q}{(2 \pi )^3} \, \frac{e^{ i \vec q \cdot \vec r}}{(\vec q \, ^2 + M_{\pi}^2 )} = 
\frac{1}{4 \pi r} e^{-M_\pi r} \,, \nn
h_2 (r ) &=& \int \frac{d^3 q}{(2 \pi )^3} \, \frac{e^{ i \vec q \cdot \vec r}}{(\vec q \, ^2 + M_{\pi}^2 )^2} = 
\frac{1}{8 \pi M_\pi} e^{-M_\pi r} \,,\nn
g (r ) &=& \int \frac{d^3 q}{(2 \pi )^3} \, e^{ i \vec q \cdot \vec r} = \delta^3 (r ) \,.
\eeqa
The isospin--violating 3NF in eqs.~(\ref{3NFisosp1}), (\ref{3NFisosp2}), (\ref{3NFisosp3})
(\ref{3NFisosp4}) and (\ref{3NFisosp5}) can now be expressed in terms of the operators $O_{ijk}^{1\ldots 5}$
defined above:
\beqa
V^{\rm 3 N} &=& \sum_{i \not= j \not= k} \, \left( \frac{g_A}{2 F_\pi } \right)^2 \Bigg\{
(\fet \tau_i \cdot \fet \tau_j ) \tau_k^3  \left[ - 2 \left( \frac{g_A}{2 F_\pi } \right)^2 \delta m \; O_{ijk}^2 -
\frac{1}{4 F_\pi^2}  (\delta m )^{\rm str.} O_{ijk}^3 + 2 \delta m \,C_T  O_{ijk}^5 \right] \nn
&&{} + (\fet \tau_i \cdot \fet \tau_k ) \tau_j^3  \left[ 2 \left( \frac{g_A}{2 F_\pi } \right)^2  \delta m  \; O_{ijk}^2 +
\frac{1}{2 F_\pi^2}  (\delta m )^{\rm str.} \, O_{ijk}^3 - 
2 \delta m \, C_T  O_{ijk}^5 \right] \nn
&& {} + [ \fet \tau_i \times \fet \tau_j ]^3 \left( \frac{g_A}{2 F_\pi } \right)^2  2 \delta m \;  O_{ijk}^1 \nn
&& {} + \tau_i^3\, [ \fet \tau_j \times \fet \tau_k ]^3 \,  \frac{1}{F_\pi^2} \, \delta M_\pi^2  \, c_4 \;  O_{ijk}^1 \nn
&& {} + \tau_i^3 \tau_j^3 \,\left[ \frac{2}{F_\pi^2} \delta M_\pi^2 c_3 O_{ijk}^2 + f_1 e^2 O_{ijk}^3
-  \frac{4}{F_\pi^2} c_1 M_\pi^2 \delta M_\pi^2  O_{ijk}^4 - \frac{1}{2 g_A} D \delta M_\pi^2 O_{ijk}^5 \right] \Bigg\}
\eeqa 
Notice that the expressions for the operators $O_{ijk}^{1\ldots 5}$ in eqs.~(\ref{o1})--(\ref{o5}) are 
singular at short distance and need to be regularized. If one chooses  to work with the local regulating functions, 
the regularized expressions can easily be obtained by an appropriate modification of the functions $h_1 (r )$,
 $h_2 (r )$ and $g (r )$.

\end{document}